\documentclass[aip,
 amsmath,amssymb,
 reprint,%
]{revtex4-2}
\pdfoutput=1 

\usepackage{graphicx}
\usepackage{dcolumn}
\usepackage{bm}

\usepackage[utf8]{inputenc}
\usepackage[T1]{fontenc}
\usepackage{mathptmx}
\usepackage{etoolbox}
\usepackage{float}
\usepackage{upgreek}
\usepackage{hyperref}
\usepackage{xurl}
\usepackage{xspace}
\newcommand{\micron}{$\upmu \mathrm{m}$\xspace}

\makeatletter
\def\@email#1#2{%
 \endgroup
 \patchcmd{\titleblock@produce}
  {\frontmatter@RRAPformat}
  {\frontmatter@RRAPformat{\produce@RRAP{*#1\href{mailto:#2}{#2}}}\frontmatter@RRAPformat}
  {}{}
}%
\makeatother
\begin{document}

\preprint{AIP/123-QED}

\title[]{Suspended Z-cut lithium niobate waveguides for stimulated Brillouin scattering}
\author{Lisa-Sophie Haerteis}
\affiliation{ 
Institute of Photonics and Advanced Sensing (IPAS), University of Adelaide, Adelaide SA 5005, Australia
}
 \email{lisa.haerteis@adelaide.edu.au}

\author{Yan Gao}
\affiliation{%
Department of Microtechnology and Nanoscience, Chalmers University of Technology, Göteborg, Sweden}%

\author{Aditya Dubey}
\affiliation{%
RMIT University, Melbourne, VIC 3001, Australia
}%

\author{Miko\l aj K. Schmidt}
\affiliation{%
School of Mathematical and Physical Sciences, Macquarie University, New South Wales 2109, Australia
}%

\author{Peter Thurgood}
\affiliation{%
RMIT University, Melbourne, VIC 3001, Australia
}%
\author{Guanghui Ren}
\affiliation{%
RMIT University, Melbourne, VIC 3001, Australia
}%

\author{Jochen Schröder}
\affiliation{%
Department of Microtechnology and Nanoscience, Chalmers University of Technology, Göteborg, Sweden}

\author{David Marpaung}%
\affiliation{ 
Nonlinear Nanophotonics Group, MESA+ Institute of Nanotechnology, University of Twente, Enschede, Netherlands}

\author{Arnan Mitchell}
\affiliation{%
RMIT University, Melbourne, VIC 3001, Australia
}%

\author{Michael J. Steel}
\affiliation{%
School of Mathematical and Physical Sciences, Macquarie University, New South Wales 2109, Australia
}%

\author{Andreas Boes}
\affiliation{ 
Institute of Photonics and Advanced Sensing (IPAS), University of Adelaide, Adelaide SA 5005, Australia
}

\date{09 April 2025}

\begin{abstract}
On-chip stimulated Brillouin scattering (SBS) has recently been demonstrated in thin-film lithium niobate (TFLN), an emerging material platform for integrated photonics offering large electro-optic and nonlinear properties. While previous work on SBS in TFLN have focused on surface SBS, in this contribution we experimentally demonstrate, for the first time, backward intra-modal SBS generation in suspended Z-cut TFLN waveguides. Our results show trapping of multiple acoustic modes in this structure, featuring a multi-peak Brillouin gain spectrum due to the excitation of higher-order acoustic modes. The findings expand the TFLN waveguide platform exploration for SBS interactions and provide a crucial step towards realizing optical processors for microwave signals or sensors integrated on TFLN.
\end{abstract}

\maketitle

\section{\label{sec:level1}Introduction}
Stimulated Brillouin scattering (SBS) is an inelastic scattering process that arises from the coherent nonlinear interaction between optical fields and GHz acoustic waves~\cite{Wolff2015,Merklein2015, JESipe_2016,Eggleton2019,Wiederhecker2019}. SBS in photonic integrated circuits has made remarkable progress over the last decade~\cite{Pant:11, Shin2013, Laer_2015, Gyger2020, Botter2022, neijts2024, ye2024brillouinphotonicsenginethinfilm, Rodrigues2025}, however, an ongoing challenge for many photonic integrated circuit platforms is ensuring the simultaneous confinement of optical and acoustic waves, since guidance of both fields by total internal reflection is frequently not possible. Nevertheless, efficient on-chip SBS has been demonstrated in various waveguide material platforms, such as chalcogenide~\cite{Pant:11}, silicon nitride~\cite{Gyger2020,Botter2022} and silicon~\cite{Shin2013,VanLaer2015}. Thin-film lithium niobate (TFLN) has recently gained attention as a platform that supports on-chip SBS as simultaneously demonstrated by Ye et al.~\cite{ye2024brillouinphotonicsenginethinfilm}, Rodrigues et al.~\cite{Rodrigues:23, Rodrigues2025} and Yang et al.~\cite{Yang2023}. Realizing efficient SBS in TFLN waveguides is attractive as electro-optic modulation and SBS light manipulation could be accommodated on a single chip to form a compact and efficient microwave photonic processor~\cite{MORRISON201485, Choudhary:16, Garrett2023}, suitable for applications in quantum photonics, sensing, and optical communications. 

While the demonstrated SBS in TFLN is promising, previous work has mainly focused on surface SBS, with a surface acoustic wave (SAW) confined at the surface of the TFLN ridge waveguide~\cite{Rodrigues2025, ye2024brillouinphotonicsenginethinfilm}. This is the case, as for the most commonly used TFLN configuration, the LN thin film is placed on top of a $\mathrm{SiO_2}$ buffer layer. In that configuration, the SAW is the only acoustic mode that is efficiently confined, as other acoustic modes leak into the underlying $\mathrm{SiO_2}$ layer due to the lower acoustic velocity of $\mathrm{SiO_2}$ compared to LN. Acoustic leakage therefore remains a problem in TFLN on $\mathrm{SiO_2}$ structures \cite{Poulton:13, Wolff:21, Rodrigues:23}, hindering the exploration of higher-order acoustic modes with various displacement fields, which occur at other frequencies, and the diverse Brillouin gain spectrum that TFLN could offer. 

The investigation of confinable acoustic modes in TFLN has the potential to contribute to a more comprehensive understanding of the range of Brillouin frequency shifts that LN offers and its suitability for SBS sensing. With regard to sensing applications, a subject of research is the generation of multi-peak Brillouin gain spectra to use the fundamental and higher-order acoustic modes for multi-parameter sensing as each acoustic mode exhibits its own sensitivity towards e.g. temperature and strain. Using these acoustic modes simultaneously is then crucial to separate between those two parameters. Currently, this approach uses complex fiber structures to generate such multi-peak Brillouin spectra~\cite{Xu:16, Sheng:24}. Investigating multi-peak SBS in TFLN waveguides might then allow replacing these complex fiber profiles with a simple and compact waveguide, making TFLN an attractive material for sensing applications.

A path towards addressing diverse Brillouin shifts in TFLN waveguides is the realization of suspended waveguides to improve acoustic confinement in such waveguides, similar to previous demonstrations in silicon~\cite{VanLaer2015, Kittlaus2016,Wang2022} and most recently in X-cut TFLN~\cite{yu2024onchipbrillouinamplifiersuspended}. Accordingly, in this contribution we investigate backward intra-modal SBS in suspended Z-cut TFLN waveguides fabricated by removing the underlying silica layer via underetching. Both numerical and experimental studies suggest the support of multiple, higher-order acoustic modes in the structure as Brillouin shifts between 8 and 10\,GHz can be observed. The results demonstrate that LN offers a variety of Brillouin frequency shifts that may be suitable for exploration in future microwave photonic and multi-peak sensing applications. 

\section{Numerical modeling} \label{sec:modeling}
\begin{figure}
\centering
\includegraphics[width=0.4\textwidth]{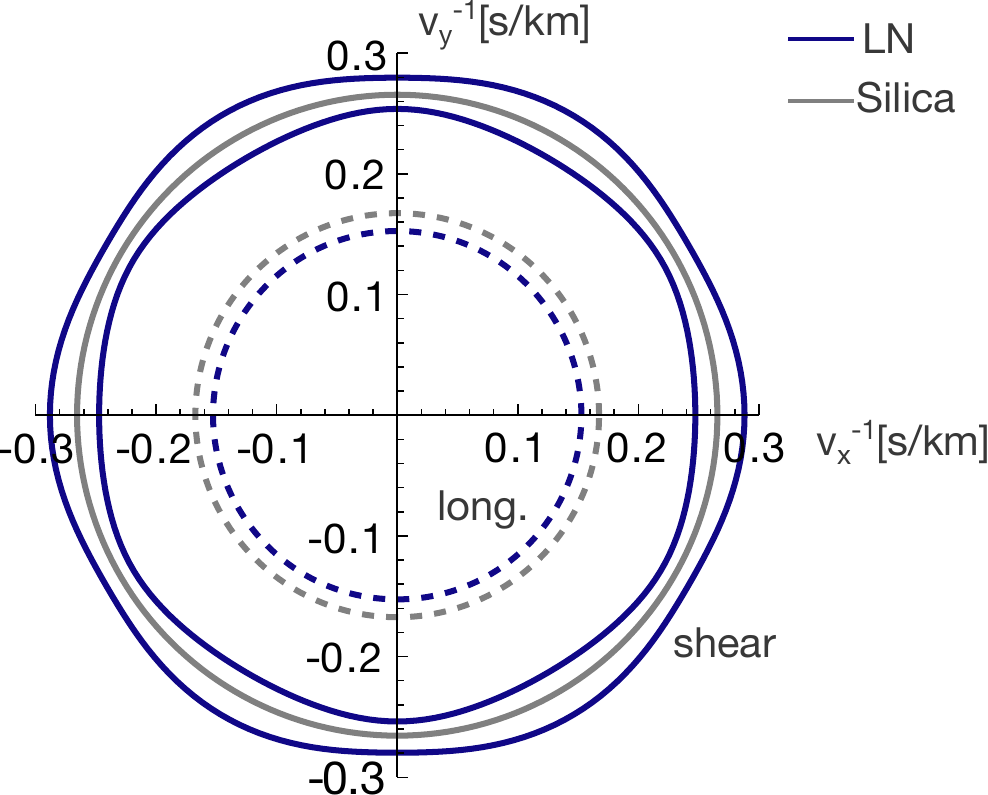}
\caption{Slowness curves (inverse velocity) of different acoustic modes in bulk Z-cut LN (blue) and bulk silica ($\mathrm{SiO_2}$, grey). The solid lines represent acoustic shear modes, while the dashed lines represent the longitudinal modes. Only two curves are visible for silica, as the two shear modes are degenerate.}
\label{fig:slowness_curve}
\end{figure}
The confinement of optical and acoustic waves in waveguides plays an important role in on-chip SBS. Optical confinement can be achieved using a high-refractive-index waveguide material to create a large index contrast between the waveguide core and the substrate/cladding. LN has an ordinary refractive index of approximately 2.237, while $\mathrm{SiO_2}$, a commonly used material underneath the waveguide (buffer layer), has a refractive index of 1.444 at a wavelength of 1.55\,\micron. This provides a sufficiently strong refractive index contrast for tight confinement of the optical field. In contrast, confining acoustic waves in a waveguide core is more challenging due to the need to confine both longitudinal and shear acoustic waves within the core. In general, shear waves travel significantly slower than longitudinal waves. In Z-cut LN the slowest longitudinal wave has a phase velocity of 6546\,m$\mathrm{s}^{-1}$, while the slowest shear wave propagates at 3575\,m$\mathrm{s}^{-1}$, requiring a large difference in velocity between core and cladding materials for effective confinement of acoustic waves~\cite{Eggleton2019,Wolff:21}. To visualize this issue, we plot the slowness (inverse velocity) curves of both Z-cut LN and $\mathrm{SiO_2}$ for longitudinal and shear waves in Fig.~\ref{fig:slowness_curve}. It is worth noting, that the acoustic velocities (along with the photoelastic coefficients) depend on the crystalline orientation as well as the waveguide orientation angle of LN \cite{LiNbO,Rodrigues:23}.
Conventional acoustic guidance would require that all grey curves ($\mathrm{SiO_2}$) lie inside all blue curves (LN), so that the core wave speed in LN is slower than all acoustic waves in $\mathrm{SiO_2}$~\cite{Wolff:21}. Since this is not the case in Fig.~\ref{fig:slowness_curve}, acoustic leakage should be expected for the standard TFLN/silica platform. Despite this, recent work~\cite{ye2024brillouinphotonicsenginethinfilm, Rodrigues:23, Rodrigues2025} successfully demonstrated SBS in Z- and X-cut TFLN waveguides on a $\mathrm{SiO_2}$ layer, by harnessing surface acoustic waves (SAW) that maintain confinement despite the presence of the $\mathrm{SiO_2}$ layer. However, the SAW is the only acoustic mode confined in the TFLN on $\mathrm{SiO_2}$ platform, with higher velocity acoustic resonances leaking into the $\mathrm{SiO_2}$ and not contributing to SBS interactions. 
\begin{figure}
\includegraphics[width=0.47\textwidth]{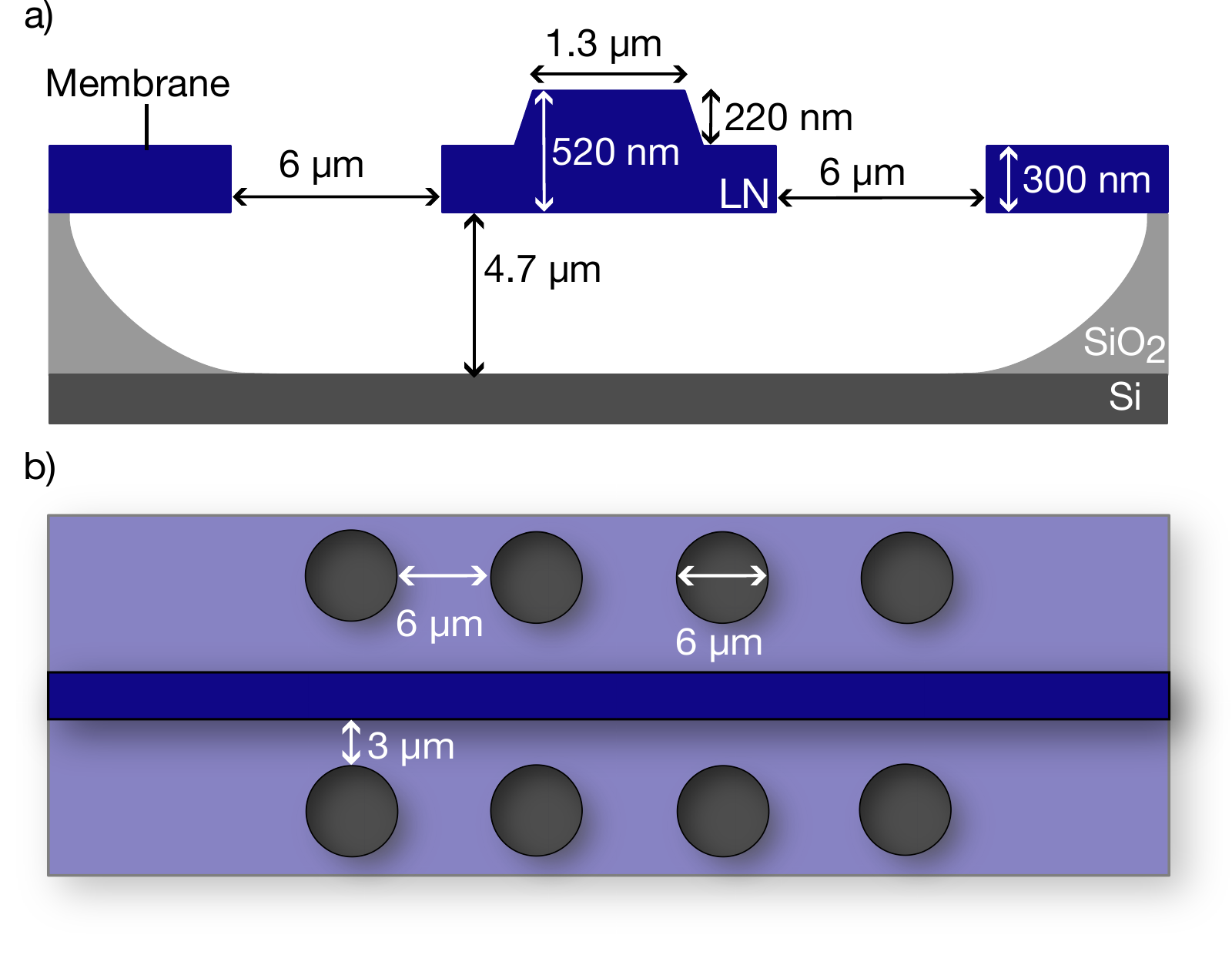}
\caption{(a) Schematic cross section of the suspended waveguide design; (b) schematic top view of the waveguide.}
\label{fig_waveguide_schem}
\end{figure}

To avoid this leakage channel, we consider a suspended Z-cut TFLN ridge waveguide structure as shown in Fig.~\ref{fig_waveguide_schem}a) and b). The large side holes are to facilitate the underetch procedure. To numerically calculate the Brillouin gain spectrum of the waveguide, we used the open-source finite element software NumBAT~\cite{sturmberg,numbatgithub}. The investigated LN ridge waveguide structure has a top width of $w=1300$\,$\mathrm{nm}$, a height of $h=520$\,$\mathrm{nm}$ and a slab (or membrane) thickness of $h_\mathrm{slab}=300$\,$\mathrm{nm}$. The ridge waveguide has a sidewall angle of $20^\circ$, due to the argon ion milling fabrication~\cite{Gao:23}. The waveguide is modeled as a floating structure with the surrounding material being vacuum to mimic a suspended waveguide. The stiffness tensor components of LN are taken from~Ref.\citenum{Rodrigues:23} and a quality factor of 90 is assumed for the simulations to match the experimental results that indicate Brillouin linewidths of approximately 100\,MHz.

\begin{figure}[ht]
\includegraphics[width=0.43\textwidth]{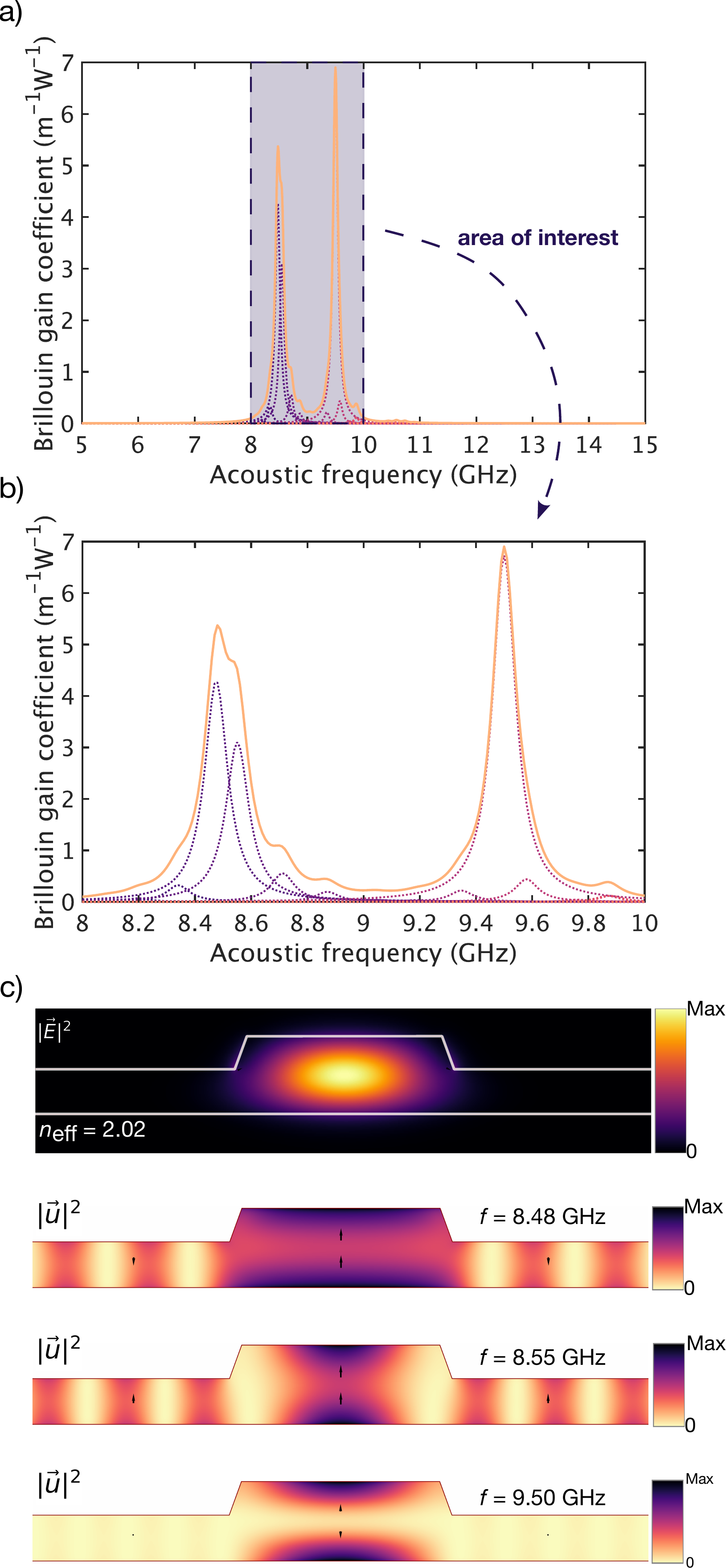}
\caption{Numerical calculations of (a) backward intra-modal Brillouin gain spectrum for the quasi-TE optical mode shown in the top panel of part (c), the dotted lines visualize individual acoustic modes (shown in the lower panels of part (c) as displacement field $|\vec u|^2$), mediating the SBS process. The solid line is the sum of all the individual contributions; (b) the zoom-in of the spectrum in (a); (c) the optical mode profile as electric field $|\vec E|^2$ for quasi-TE mode and mechanical mode profiles with acoustic displacement field $|\vec u|^2$.}
\label{fig:spectrum}
\end{figure}
In Fig.~\ref{fig:spectrum}a) we plot the calculated Brillouin gain spectrum for backward intra-modal SBS with the fundamental optical quasi-TE mode at 1.55\,\micron\,(see top image in Fig.~\ref{fig:spectrum}c). A zoomed-in plot in Fig.~\ref{fig:spectrum}b) shows two features, centered at 8.49 GHz and 9.50 GHz. Since the NumBAT solver calculates gain mediated by each acoustic mode (with contributions marked with dotted line), we can readily identify that the lower frequency feature is due to two modes at 8.48 GHz and 8.55 GHz. To further analyze the numerical Brillouin gain spectrum, Fig.~\ref{fig:spectrum}c) illustrates the respective electric field $|\vec E|^2$ of the optical mode (quasi-TE mode) and the displacement fields $|\vec u|^2$ of the acoustic modes, showing the two acoustic modes at Brillouin shifts of 8.48\,GHz and 8.55\,GHz as well as the higher frequency one at 9.50\,GHz. The acoustic modes are mostly localized on the top and bottom of the waveguide and are predominantly shear polarized, with transverse fractions ranging between 0.89 and 0.96 (the transverse polarization fraction is defined in \ref{app:modecalc}). The model predicts the highest gain coefficient for the acoustic mode around 9.50\,GHz with a slightly higher amplitude than the other two resonances. 

To conclude, our numerical calculations indicate that suspended TFLN can support richer SBS gain profiles compared to the SAW-only Brillouin spectra experimentally reported for TFLN so far~\cite{ye2024brillouinphotonicsenginethinfilm, Rodrigues:23, Yang2023}.
\begin{figure}
\includegraphics[width=0.425\textwidth]{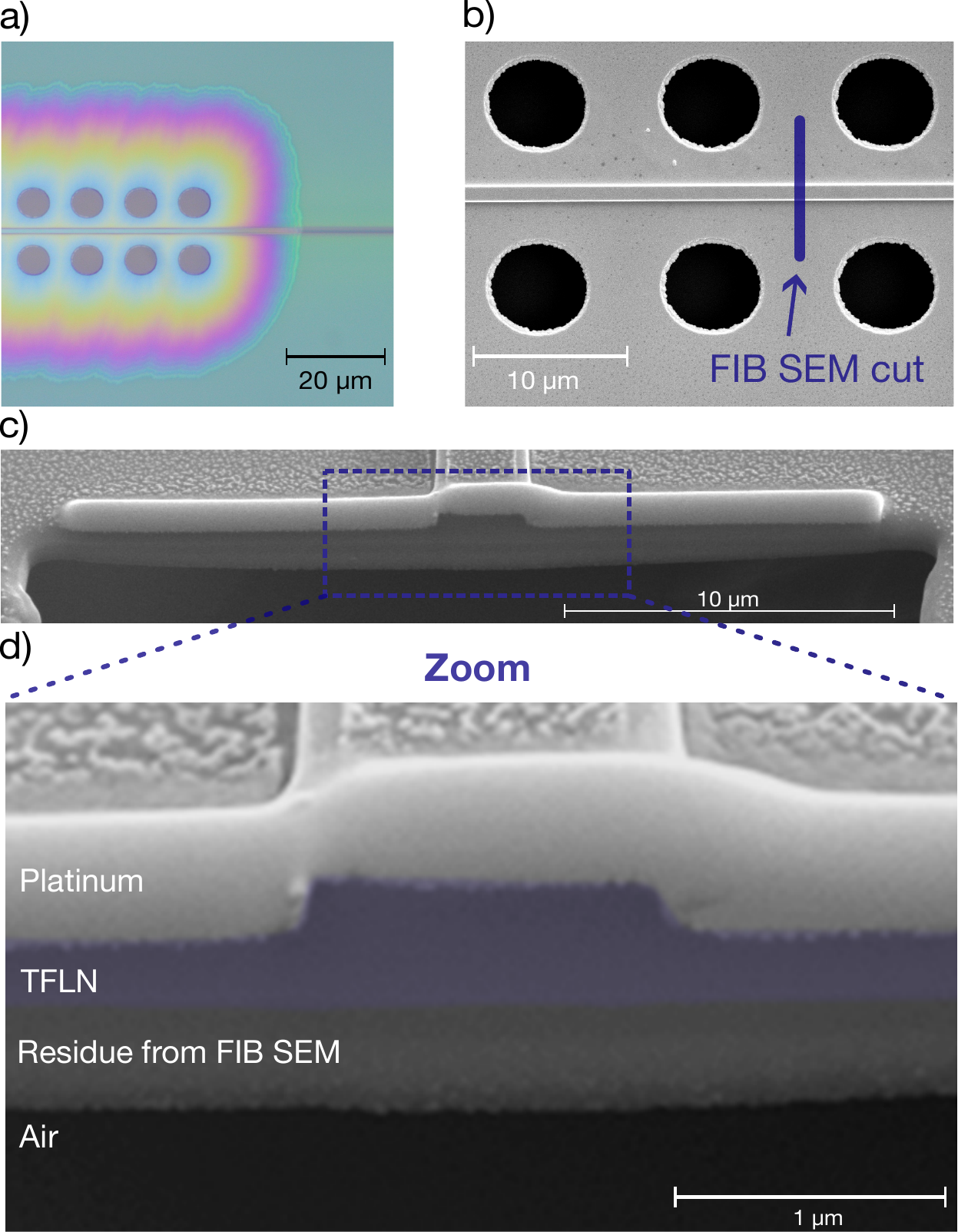}
\caption{(a) Bright-field microscope image of the suspended waveguide; (b) SEM image of the waveguide's top view; (c) cross section FIB SEM image of the suspended waveguide. The cut for the FIB SEM is indicated in (b); (d) Zoom into the cross section of the suspended waveguide, indicating the different material layers.}
\label{fig:wguideSEM}
\end{figure}
\section{Experimental methodology}
\subsection{Waveguide design and fabrication}
The initial waveguide is fabricated on a 520\,nm thick Z-cut TFLN wafer (NANOLN). The fabrication process~\cite{Gao:23} starts by preparing the sample via solvent and standard cleaning, followed by a MaN 2405 resist spin-coating on the sample. The waveguide pattern is then defined on the resist via 100-kV e-beam exposure (Raith EBPG 5200), using multipass exposure to reduce the waveguide sidewall roughness. The sample is then dry-etched using a reactive ion beam etching tool (RIE tool, Oxford Ionfab 300 Plus) with Ar plasma. Subsequent solvent and SC1 cleaning steps effectively remove the resist. SC1 cleaning after etching serves a dual purpose: it removes by-products generated during ion beam etching and may also slightly etch the LN waveguide sidewalls, potentially improving their surface roughness~\cite{Gao:23}. 

Next, we prepare the sample for underetching by etching periodic holes on either side of the ridge waveguide using photon lithography (MLA150, Heidelberg Instruments) with bilayer resists LOR3A/S1805, followed then by another dry etching step as described above. The resulting structure is illustrated in Fig.~\ref{fig_waveguide_schem}b) and \ref{fig:wguideSEM}b). The distance between holes and waveguide (edge to edge) is 3\,\micron while the distance between holes is 6\,\micron, a design tailored to create a stable membrane withholding the underetching procedure. The holes are fully etched (through the TFLN). Subsequently, the waveguide is suspended using a wet etching method, immersing the chip in a 49\% HF solution for 5 minutes. The 49\% HF solution removes the underlying silica through the fully etched holes, creating a suspended waveguide held by the membrane surrounding the TFLN waveguides. The length of the suspended waveguide is 1.1\,cm. Top-down views from optical microscopy and SEM are shown in Fig.~\ref{fig:wguideSEM}a) and b). Figure \ref{fig:wguideSEM}a) conveniently illustrates the removal of silica where the different colors indicate the different thicknesses of the buffer layer. A FIB SEM image of the suspended waveguide is given in Fig.~\ref{fig:wguideSEM}c) with a zoom into the cross section in Fig.~\ref{fig:wguideSEM}d), confirming the successful suspension of the waveguide.

\subsection{Experimental characterization}
\begin{figure}
\includegraphics[width=0.45\textwidth]{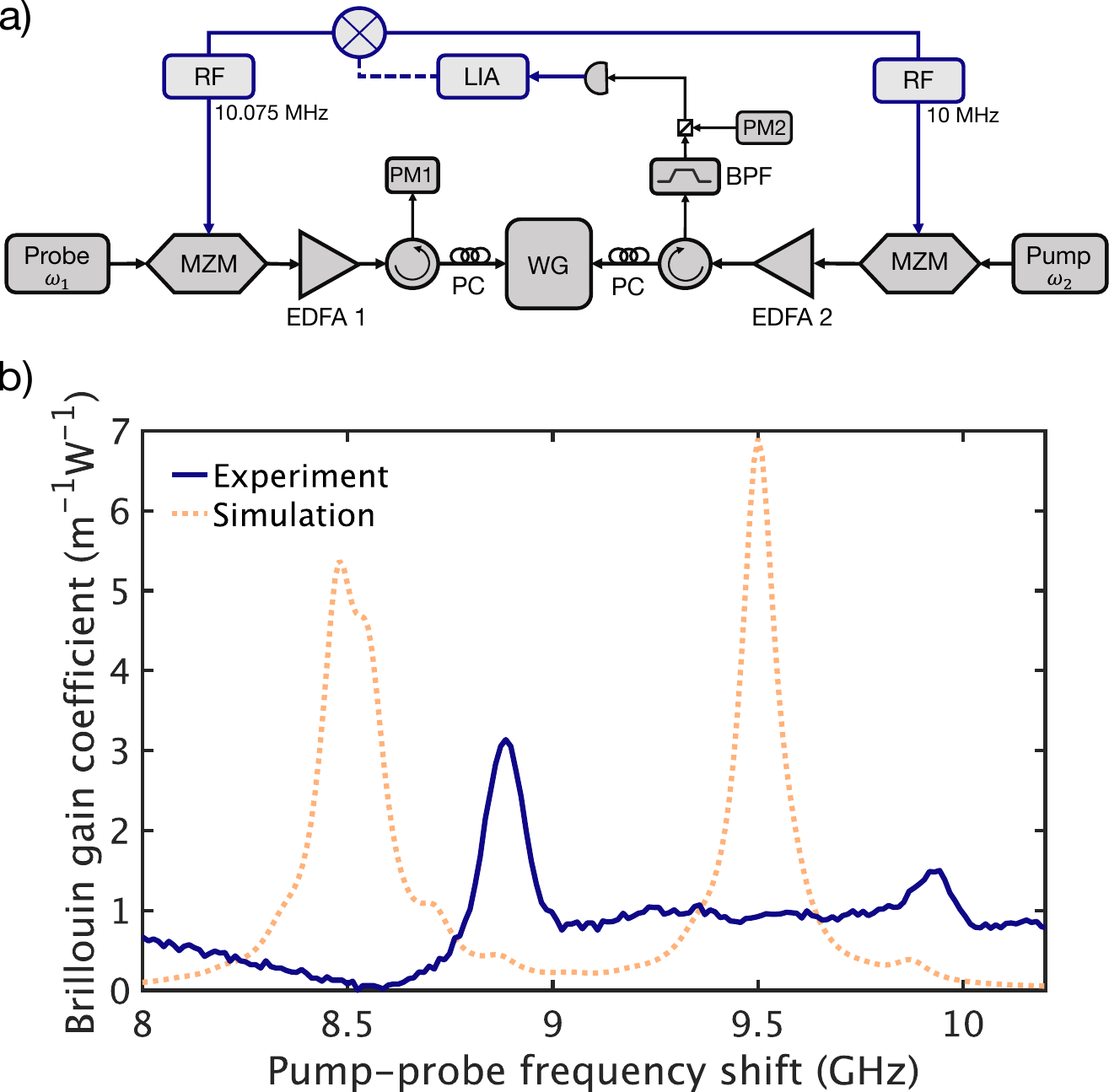}
\caption{(a) Double intensity-modulated pump-probe measurement technique for Brillouin gain response characterization, PM: power meter, LIA: lock-in amplifier, RF: radio frequency signal, MZM: Mach-Zehnder; (b) numerical and experimental Brillouin gain spectrum of the suspended waveguide.}
\label{fig:expt}
\end{figure}

For Brillouin gain characterization of the suspended waveguide we use a double intensity-modulated pump-probe lock-in amplifier measurement technique~\cite{Botter2022} as schematically depicted in Fig.~\ref{fig:expt}a). The pump signal is delivered by a DFB laser diode (Thorlabs DFB1550P) at 1552\,nm and a tunable laser source (Toptica DLC pro with CTL) operating around 1552\,nm is used as the probe signal. Pump and probe are coupled into the waveguide from opposite directions and are both intensity-modulated with Mach-Zehnder modulators (MZM, JDSU 10G Mod) fed by 10\,MHz and 10.075\,MHz sinusoidal signals, respectively. The pump signal is amplified to 2\,W, while the probe signal is amplified to 144\,mW using erbium-doped fiber amplifiers (EDFA, Amonics AEDFA-33-B-FA and AEDFA-PA-35-B-FA, respectively). The probe is swept during the measurement for a pump-probe detuning between 0 and 12\,GHz as we expect the Brillouin shifts of the waveguides to lie between 7 and 10\,GHz. The transmitted probe is measured with a photo detector (PD, Optilab PD-23-C-AC) and then processed by a lock-in amplifier (LIA, Stanford Research Systems SRS830 DSP), referenced at 75 kHz.

\subsection{Experimental Results and Discussion}
The SBS measurement was conducted with a pump power of 1.6\,W and taking into account facet losses of 6\,dB per facet, this resulted in an on-chip pump power of approximately 400\,mW. The measured backward intra-modal Brillouin gain spectrum is shown in Fig.~\ref{fig:expt}b), where the Brillouin gain coefficient is extracted by comparing the amplitude of a resonance to the fiber SBS resonance measured simultaneously with the on-chip SBS~\cite{Botter2022}. The Brillouin gain characterization reveals that the waveguide yields two resonances at 8.88\,GHz and 9.92\,GHz as depicted in Fig.~\ref{fig:expt}b). A Lorentzian fit was used to predict the linewidth of the Brillouin peaks. The linewidth of the resonance at 8.88\,GHz is 114\,MHz, while the second resonance exhibits a linewidth of 62\,MHz. Furthermore, the resonance at 8.88\,GHz exhibits a higher gain coefficient than the second resonance.

In comparison with the modeling, the experimental resonances exhibit an upwards frequency shift of about 0.4\,GHz, while the separation \textit{between} the resonances is 1.04\,GHz, close to the calculated value of 1.1\,GHz. We also observe that the first peak exhibits a linewidth larger than that of the second peak for both the modeling and the experimental results. This can be explained by the resonance at lower frequencies being mediated by two narrowly spaced acoustic modes as predicted by the model. It is also notable that the second resonance at 9.92\,GHz is much lower in efficiency according to the experimental data, when compared to the modeling.
While overall the model and experiment agree well, the mismatch in Brillouin frequencies from model and experiment may result from discrepancies in the geometrical parameters and simulation assumptions. Appendix~\ref{app:parametricsweep} explores the dependence of the Brillouin shift on several waveguide dimensions, and reveals that the frequency shift and separation is a quite sensitive function of the rib thickness, and to a lesser extent, of the membrane thickness. A variation of the geometrical parameters and shape of the waveguide from our modeled structure may occur due to fabrication tolerances. For example the TFLN may be slightly etched during the wet etching process since the -Z-face of LN is facing downward, ultimately being exposed to the etching solution as soon as the silica is removed underneath \cite{sones}. Additionally, a slight sagging of the waveguide structure due to the suspension and consequent internal strains might affect the Brillouin frequencies. This can be avoided in future designs by carefully choosing the membrane thickness and optimizing the underetching procedure.

\section{Conclusion}
This work presents the first experimental demonstration of stimulated Brillouin scattering in suspended Z-cut TFLN waveguides. The waveguide structure was designed to explore the acoustic confinement of multiple modes in TFLN-based waveguides for efficient SBS interaction. This behavior was confirmed with an experimental demonstration of backward intra-modal SBS in the suspended waveguides using a double intensity-modulated pump-probe technique. Two resonances were found in the experimental data with Brillouin shifts of 8.88\,GHz and 9.92\,GHz, separated by 1.04\,GHz. The highest experimentally demonstrated Brillouin gain coefficient in our measurements was $3.14$\,$\mathrm{m}^{-1}\mathrm{W}^{-1}$ mediated by the acoustic mode with frequency 8.88\,GHz. The gain coefficient can be further enhanced through optimization of the waveguide geometry and fabrication procedure.

These findings expand our understanding of the acoustic properties of the material and show that TFLN supports a rich variety of acoustic modes mediating SBS. This multi-peak Brillouin characteristics in TFLN are particularly attractive for the on-chip integration of sensing applications as recently demonstrated in complex fiber structures \cite{Xu:16, Sheng:24}.The exact Brillouin frequencies and their separation can be tailored with careful waveguide design as shown in Appendix~\ref{app:parametricsweep}. 

\begin{acknowledgments}
This work has been supported through the Australian Research Council (grant no. DP220100488, DE230100964 and DE220101272) and the Swedish Research Council (grant no. VR-2021-04241). The authors would like to express their gratitude towards Aminesh Basak and Ken Neubauer at Adelaide Microscopy for their support with SEM imaging. The initial LN samples were fabricated at Myfab Chalmers. The authors would also like to thank Dashen Dong for his assistance with the underetching procedure.
\end{acknowledgments}

\section*{Author declarations}
\subsection*{Conflict of Interest}
The authors have no conflicts to disclose.

\subsection*{Author contributions}
\textbf{Lisa-Sophie Haerteis}: Conceptualization (equal); Data curation (equal); Formal analysis (equal); Investigation (equal); Methodology (equal); Software(equal); Visualization (equal); Writing - original draft (equal); Writing - Review \& Editing (equal). \textbf{Yan Gao}: Resources (equal); Writing - original draft (supporting). \textbf{Aditya Dubey}: Resources (equal); \textbf{Miko\l aj K. Schmidt}: Conceptualization (equal); Formal analysis (equal); Investigation (supporting); Software (supporting); Writing - Review \& Editing (equal). \textbf{Peter Thurgood}: Resources (equal); \textbf{Guanghui Ren}: Resources (equal);  \textbf{Jochen Schröder}: Resources (supporting); Writing - Review \& Editing (equal). \textbf{David Marpaung}: Funding Acquisition (equal); Writing - Review \& Editing (equal). \textbf{Arnan Mitchell}: Resources (supporting); \textbf{Michael J. Steel}: Conceptualization (equal); Formal analysis (equal); Funding acquisition (equal); Investigation (equal); Software (equal); Supervision (equal); Writing - original draft (equal); Writing - Review \& Editing (equal). \textbf{Andreas Boes}: Conceptualization (equal); Funding acquisition (equal); Investigation (equal); Methodology (equal); Project administration (equal); Resources (equal); Supervision (lead); Writing - original draft (equal); Writing - Review \& Editing (equal).

\section*{Data availability}
The data that support the findings of this study are available from the corresponding author upon reasonable request.

\appendix

\section{Simulation details}

\subsection{Material properties}
We model the waveguide with the \textit{trapezoidal\_rib} structure provided in NumBAT. The values for the photo-elastic and stiffness tensors are taken from published work~\cite{Rodrigues:23} and are summarized in the material file \textit{LiNbO3\_2023\_Rodrigues.json} which can be found at \url{https://github.com/michaeljsteel/NumBAT/tree/master/backend/material\_data}. The material was rotated to ``z-cut'' crystal orientation. The suspended waveguide is modeled as being entirely surrounded by vacuum.

\subsection{Meshing details}
The computational domain was discretized using triangular mesh elements. The final elastic mesh consists of 10370 elements in total with 21343 nodes. Mesh refinement is applied in the waveguide core and slab to capture the optical mode field and acoustic displacement fields. The mesh element size inside the waveguide is about 20\,nm with increasing element size away from the waveguide boundaries. 

In terms of the NumBAT implementation, the FEM mesh is created with the open-source program Gmsh~\cite{sturmberg} and is defined by three parameters, specifically $lc_\mathrm{bkg}, lc_\mathrm{refine,1}$ and $lc_\mathrm{refine,2}$. We set the background mesh size to $lc_\mathrm{bkg}=.08$ resulting in 14 mesh points on the outer boundary. The next parameter $lc_\mathrm{refine_1}$ helps to refine the mesh at the interface between materials, where a larger value gives a finer mesh. We set the value to $lc_\mathrm{refine,1}=30.0$. To refine the mesh in the center of the waveguide, $lc_\mathrm{refine,2}$ can be used. We choose a relatively high value of 20 to ensure a fine mesh. 

\numberwithin{equation}{section} 
\subsection{Mode calculations} \label{app:modecalc}
The transverse fraction $f_t$ of the displacement field $\vec u(x,y)$ of an acoustic mode, as mentioned in section \ref{sec:modeling}, is defined as follows:
\begin{equation}
    f_t = \frac{f_x + f_y}{f_x + f_y+f_z} 
\end{equation}
where the individual component values $f_i$ for $i=x,y,z$ satisfy
\begin{equation}
    f_i = \iint |\vec{u}_i|^2 \, \mathrm{d}x \mathrm{d}y,  
\end{equation}

\section{Parametric study}\label{app:parametricsweep}

\begin{figure*}
\centering
\includegraphics[width=0.95\textwidth]{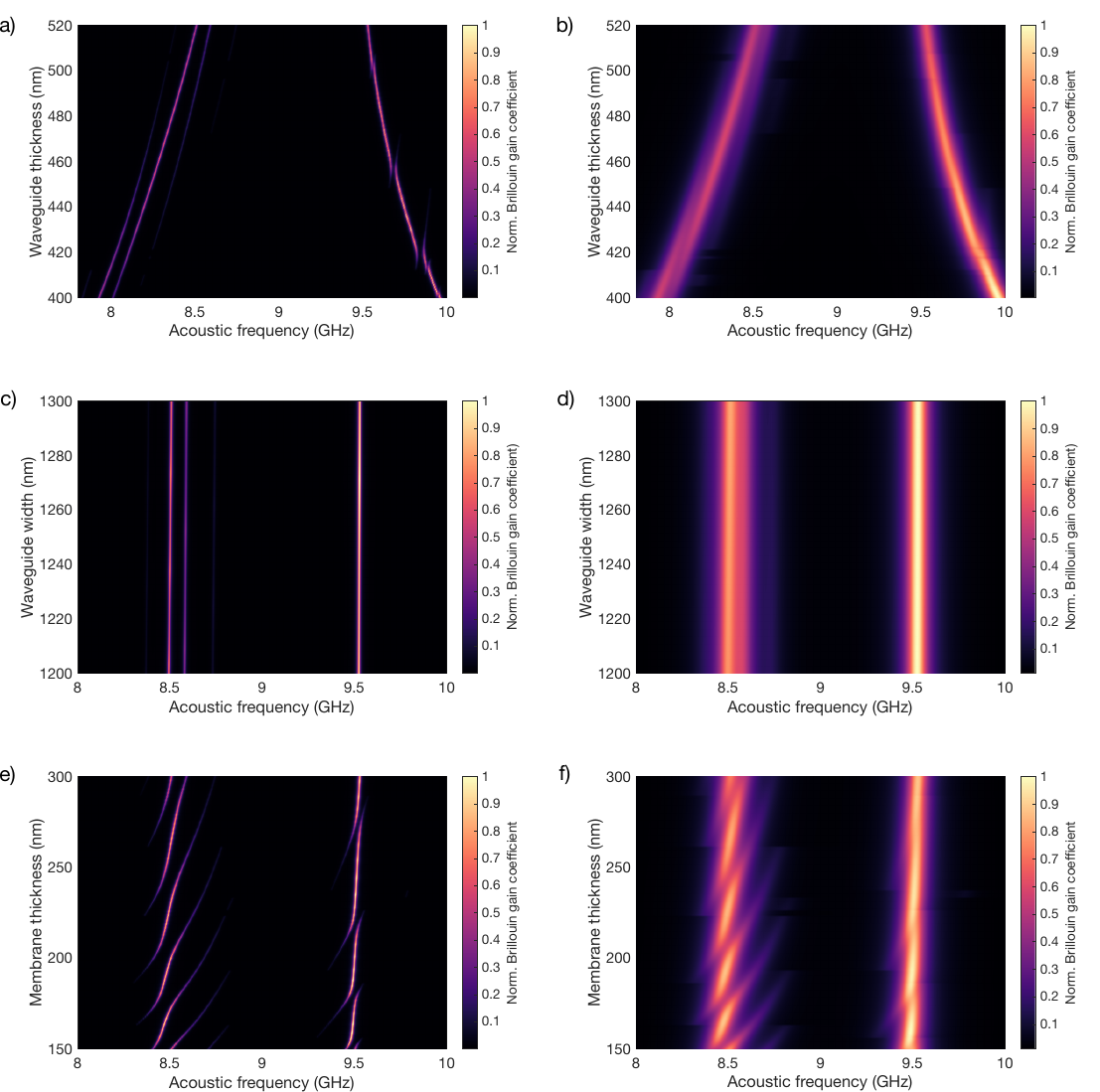}
\caption{Parametric study. The graphs illustrate the impact of waveguide thickness, width, and membrane thickness variations. Simulations for $Q=900$ are shown on the left, and $Q=90$ on the right: (a) Brillouin shift evolution as a function of waveguide thickness with $Q=900$. The waveguide width and membrane thickness are set to 1300\,nm and 300\,nm, respectively; (b) Brillouin shift evolution as a function of waveguide thickness with $Q=90$. The waveguide width and membrane thickness are set to 1300\,nm and 300\,nm, respectively; (c) Brillouin shift evolution as a function of waveguide width with $Q=900$. The waveguide and membrane thickness are set to 520\,nm and 300\,nm, respectively; (d) Brillouin shift evolution as a function of waveguide width with $Q=90$. The waveguide and membrane thickness are set to 520\,nm and 300\,nm, respectively; (e) Brillouin shift evolution as a function of membrane thickness with $Q=900$. The waveguide width and thickness are set to 1300\,nm and 520\,nm, respectively; (f) Brillouin shift evolution as a function of membrane thickness with $Q=90$. The waveguide width and thickness are set to 1300\,nm and 520\,nm, respectively.}
\label{fig:parametric}
\end{figure*}

We perform dimensional sweeps for waveguide width, waveguide thickness and membrane thickness to evaluate the shift of the acoustic frequencies of the involved acoustic modes as a function of these dimension parameters. 
Figures~\ref{fig:parametric}a) and b) show the Brillouin gain spectrum as a function of the waveguide thickness, where we sweep the waveguide thickness between 400\,nm and 520\,nm, with $Q=900$ and $Q=90$, respectively. The involved acoustic modes around 8\,GHz tend to increase in frequency with a thicker waveguide. The acoustic frequencies range between 7.8\,GHz and 8.7\,GHz, i.e. an overall shift of about 10\%. The higher-order acoustic mode tends to decrease in frequency and shifts from around 10\,GHz to 9.5\,GHz with increasing waveguide thickness. Concluding these observations, the acoustic frequencies of the involved acoustic modes approach each other slightly over the course of this parametric sweep. 
Figures~\ref{fig:parametric}c) and d) show the evolution of the Brillouin shifts as a function of waveguide width where the width is swept between 1200\,nm and 1300\,nm, with $Q=900$ and $Q=90$, respectively. As can be seen, the acoustic frequencies shift marginally, i.e. that a width variation due to fabrication has hardly any influence on the Brillouin gain spectrum. 
The acoustic frequencies in Fig.~\ref{fig:parametric}e) and f) show a drift in frequency with increasing membrane thickness for a set waveguide width and thickness, for $Q=900$ and $Q=90$, respectively. The overall shift for the higher frequency acoustic mode is rather small, the Brillouin shift ranges between about 9.3\,GHz and 9.6\,GHz, a variation of less than 3\%. The lower frequency acoustic modes drift on a slightly larger scale, however, the offset is within 600\,MHz, which is, relatively speaking, less than 10\%. 

According to the parametric study in this section, the waveguide thickness exhibits the most significant impact, particularly on the spacing between the Brillouin shifts. The spacing between the Brillouin frequencies matches well in our prediction and experiment, indicating that our estimation of the waveguide thickness is reasonably close. To a lesser extent, the membrane thickness affects the Brillouin frequencies and could, for some part, account for the 400\,MHz mismatch between our prediction and experiment. However, additional effects occur, such as sagging of the waveguide due to the suspension and consequent internal strains, which are not accounted for in the modeling and will additionally influence the Brillouin shifts. The exact result is thus difficult to predict.

\section*{References}
\nocite{*}
\bibliography{references.bib}

\end{document}